\documentclass[epsfig,float,aps,superscriptaddress]{revtex4}
\usepackage{amsmath,bm}
\usepackage{epsfig}
\usepackage{graphicx}
\usepackage{booktabs}
\usepackage{pgf,pgfarrows,pgfnodes,xcolor}
\usepackage{tikz}
\usetikzlibrary{shapes,arrows}
\usetikzlibrary{decorations.pathreplacing}

\begin{document}
\title{PACIAE 2.0: An updated parton and hadron cascade model (program)
       for the relativistic nuclear collisions}
\author{Ben-Hao Sa}
\email{sabh@ciae.ac.cn}
\affiliation{China Institute of Atomic Energy, P. O. Box 275 (18), Beijing,
102413 China}
\affiliation{Institute of Particle Physics, Huazhong Normal University,
Wuhan, 430082 China}
\affiliation{CCAST (World Lab.), P. O. Box 8730 Beijing, 100080 China}
\author{Dai-Mei Zhou}
\affiliation{Institute of Particle Physics, Huazhong Normal University,
Wuhan, 430082 China}
\email{zhoudm@phy.ccnu.edu.cn}
\author{Yu-Liang Yan}
\affiliation{China Institute of Atomic Energy, P. O. Box 275 (18), Beijing,
102413 China}
\author{Xiao-Mei Li}
\affiliation{China Institute of Atomic Energy, P. O. Box 275 (18), Beijing,
102413 China}
\author{Sheng-Qin Feng}
\affiliation{Department of Physics, College of Science, Three Gorges
University, Yichang, 443002 China}
\author{Bao-Guo Dong}
\affiliation{China Institute of Atomic Energy, P. O. Box 275 (18), Beijing,
102413 China}
\author{Xu Cai}
\affiliation{Institute of Particle Physics, Huazhong Normal University,
Wuhan, 430082 China}
\begin{abstract}
We have updated the parton and hadron cascade model PACIAE for the
relativistic nuclear collisions, from based on JETSET 6.4 and PYTHIA
5.7 to based on PYTHIA 6.4, and renamed as PACIAE 2.0. The main
physics concerning the stages of the parton initiation, parton
rescattering, hadronization, and hadron rescattering were discussed.
The structures of the programs were briefly explained. In addition,
some calculated examples were compared with the experimental data. It
turns out that this model (program) works well.
\end{abstract}
\maketitle

\begin{flushleft}\bf\large{PROGRAM SUMMARY}\end{flushleft}

\small{
{\it Title of program:} PACIAE version 2.0

{\it Catalogue number:} ADBS

{\it Program obtained from:} CPC Program Library, Queen's University, Belfast,
N. Ireland; \\  or from yanyl@ciae.ac.cn, zhoudm@phy.ccnu.edu.cn,
sabh@ciae.ac.cn

{\it Computer for which the program is designed and others on which it has
been tested:} DELL Studio XPS and \\  others with a FORTRAN 77 or GFORTRAN
compiler

{\it Computer:} DELL Studio XPS

{\it Programming languages:} FORTRAN 77

{\it Memory required to execute with typical data:} $\approx$1G words

{\it No. of bits in a word:} 64

{\it Peripherals used:} terminal for input, terminal or printer for output

{\it No. of lines in distributed program, including test data, etc.:}
$\approx$200000

{\it Distribution format:} tar.gz

{\it Nature of physical problem:} The Monte Carlo simulation of
hadron transport (cascade) model is successful in studying the
observables at final state in the relativistic nuclear collisions.
However the high p$_T$ suppression, the jet quenching (energy loss),
and the eccentricity scaling of $v_2$ etc., observed in high energy
nuclear collisions, indicates the important effect of the initial
partonic state on the final hadronic state. Therefore the better
parton and hadron transport (cascade) models for the relativistic
nuclear collisions are highly required.

{\it Method of solution:} The parton and hadron cascade model PACIAE
is originally based on the JETSET 7.4 and PYTHIA 5.7. The PYTHIA
model has been updated to PYTHIA 6.4 with the additions of new
physics, the improvements in existing physics, and the embedding of the
JETSET model etc.. Therefore we update the PACIAE model to the
new version of PACIAE 2.0 based on the PYTHIA 6.4 in this paper. In
addition, some improvements in physics have been introduced in this
new version.

{\it Restrictions:} Depend on the problem studied.

{\it Typical running time:} Depend on the type of collision and
energy. Examples running on the DELL Studio \\  XPS are follows:
 \begin{itemize}
 \item Running 1000 events for inelastic pp collisions at $\sqrt{s}$=200 GeV
by program PACIAE 2.0a to reproduce PHOBOS data of rapidity density at
mid-rapidity, $dN_{ch}/dy$=2.25$^{+0.37}_{-0.30}$ \cite{phob2}, spends
$\approx$ 3 minutes.
 \item Running 0-6\% most central Au+Au collision at $\sqrt{s_{NN}}$=200 GeV
by program PACIAE 2.0b and PACIAE 2.0c to reproduce PHOBOS data of charged
multiplicity of 5060 \cite{phob} spends $\approx$13 seconds/event and
$\approx$265 seconds/event, respectively.
 \end{itemize}


}
\newpage
\begin{flushleft}\bf\large{LONG WRITE-UP}\end{flushleft}

\normalsize{
\section{Introduction}
The hadron transport (cascade) model is successful in describing the
relativistic nuclear collisions. However, many new phenomena
observed in the relativistic nucleus-nucleus (including
proton-nucleus) collisions at RHIC energies, such as high $p_T$
suppression \cite{star1,phen}, jet quenching (energy loss)
\cite{wang}, and elliptic flow eccentricity scaling \cite{lace}
etc., strongly indicate the important effect of the partonic initial
state on the hadronic final state. The better parton and hadron
transport (cascade) model is urgently required.

The first  parton and hadron cascade model \cite{mull} encouraged
copious publications of the similar models, such as AMPT (string
melting) \cite{ko}, PACIAE \cite{sa}, BAMPS (parton cascade only)
\cite{zhe}, PHSD \cite{cass}, and MARTINI \cite{san} etc.. Among
them the PACIAE model, a parton and hadron cascade model for
relativistic nuclear collision, is based on PYTHIA 5.7 and JETSET
7.4 \cite{sjo}. We have employed this PACIAE model successfully
studying \cite{sa4}:
\begin{itemize}
\item limiting fragmentation phenomenon in elementary and heavy-ion
collisions,
\item fluctuations and correlations in particle production,
\item direct photon production in pp and Au+Au collisions at RHIC energies,
\item property of QCD matter at RHIC and LHC energies,
\item charged particle and strange particle productions in pp collisions at 
LHC energies,
\item charged particle elliptic flow in pp collisions at LHC energies.
\end{itemize}
In this paper, the PACIAE model is updated to PACIAE 2.0 based on PYTHIA
6.4 \cite{sjo1} with the additions of new physics and the improvements in
existing physics. As the latest version PYTHIA 8.1 \cite{sjo2} ``does not
yet in every respect replaces the old code" and the physical differences
between PYTHIA 8.1 and PYTHIA 6.4 are not important for the investigation
of nucleus-nucleus collisions, so we left the connection with
PYTHIA 8.1 in the future.

PYTHIA 6.4 is a Monte Carlo event generator for relativistic
hadron-hadron collisions in hadronic level. In this model a pp
(hadron-hadron, hh) collision is decomposed into parton-parton
collisions. A hard parton-parton collision is described by the
lowest leading order perturbative QCD (LO-pQCD). The soft
parton-parton collision, non-perturbative phenomenon, is considered
empirically. Because the initial- and final-state QCD radiations and
multiparton interactions are considered in the parton-parton
scattering, the consequence of a hh collision is a parton multijet
configuration composed of di-quarks (anti-diquarks), quarks
(anti-quarks), and gluons, besides a few hadronic remnants. This
parton multijet configuration is followed by the string construction
and fragmentation (hadronization). Therefore one obtains a hadronic
final state for a hh (pp) collision. If one switches off the string
fragmentation and breaks up the di-quarks (anti-diquarks) to quarks
(anti-quarks), one then obtains a state composed of the quarks,
anti-quarks, and the gluons. This is the key point in the creation
of the parton and hadron cascade model PACIAE for the relativistic
nuclear collisions.

The PACIAE model composes of four stages of the parton initiation,
parton rescattering, hadronization, and hadron rescattering. PACIAE
2.0 has three versions: PACIAE 2.0a describing the relativistic pp
collision ($\rm{\bar pp}$, or e$^+$e$^-$, denoted as elementary collision
later) and PACIAE 2.0b as well as PACIAE 2.0c describing the relativistic
nucleus-nucleus (A+B, including p+A) collisions. The hybrid system
of units, based on the mass in GeV, length in fm, and time in fm
with $c$=1 and $hc$=1.24 GeV$\cdot$fm, is used in PACIAE 2.0.

The rest of this paper is organized as follows: Section II discussed
the main physics in the PACIAE model except the ones in PYTHIA:
\begin{itemize}
\item nucleon initiation in the position and momentum phase spaces,
\item particle propagation (cascade) and the collision criteria,
\item cross sections,
\item determination of the scattered particles,
\item diquark break-up,
\item reduction of the strange (heavy) quark suppression,
\item deexcitation of the energetic quark (antiquark) in the coalescence model.
\end{itemize}
We explain the structures of programs in section III. The examples
are calculated and compared with the experimental data in section
IV. We give the users' guide in the appendix.
\section{Physics concerned}
The PYTHIA 6.4 model involves abundant particle and nuclear physics
contents which have been discussed in \cite{sjo1} in the detail. Here we
just introduce the extra physics concerned with the nuclear
initiation, the parton rescattering, and the hadron rescattering.
 \subsection{Nucleon initiation in the position and momentum phase spaces}
  \subsubsection{Impact parameter and its sampling}
In the experiments the centrality bins are defined by the cuts in
the particle multiplicity distribution and indicated by the
percentages, $g$, in the geometric (total) cross section. However,
the centrality is conveniently defined theoretically by the impact
parameter $b$. A mapping relation between $g$ and $b$
  \begin{equation}
  b=\sqrt{g}b_{max}, b_{max}=R_A+R_B
  \end{equation}
was introduced according to the definition of geometric cross section
\cite{sa2}.

As the experiments of nucleus-nucleus collisions are always
performed in a given centrality bin the relevant theoretical
calculations should span the corresponding impact parameter interval
from $b_d$ to $b_u$. We sample impact parameter $b_i$ randomly by
  \begin{equation}
  b_i=\sqrt{\xi\times(b_u^2-b_d^2)+b_d^2},
  \end{equation}
and generate a nucleus-nucleus collision event. In the above
equation $\xi$ refers to a random number. The observable (operator)
averaged over the events generated by different $b_i$ (i=1,2,...)
can compare with the corresponding experimental data.

One may also use the systematic sampling method:
  \begin{equation}
  \int_{b_0=b_d}^{b_1}bdb=\int_{b_1}^{b_2}bdb=...=\int_{b_{n-1}}^{b_n=b_u}bdb
       =\frac{b_u^2-b_d^2}{2n},
  \end{equation}
  \begin{equation}
  b_j=\sqrt{j\frac{b_u^2-b_d^2}{n}+b_d^2} , \hspace{1cm} j=0,1,2,...,n
     ~(n~even)
  \end{equation}
The set of $b_j (j=1,3,5,...,n-1)$ is then the sampled spectral
values. We generate events by these spectral values repeatedly and compare 
the event averaged observable with the corresponding experimental data.
  \subsubsection{Geometric overlap zone and number of participant nucleons}
For a nucleus-nucleus collision A+B at a given impact parameter $b$, a
geometric overlap zone is formed between colliding nuclei in the position
phase space. The nucleons located inside this zone are the participant 
nucleons, otherwise the spectator nucleons. The geometric number of 
participant nucleons is calculated \cite{sa2} by
  \begin{align}
  N_{part}(b)=&N_{part}^A(b)+N_{part}^B(b), \\
  N_{part}^A(b)=&\rho_A\int dV\theta(R_A-(x^2+(b-y)^2+z^2)^{1/2})
                \theta(R_B-(x^2+y^2)^{1/2}), \\
  N_{part}^B(b)=&\rho_B\int dV\theta(R_B-(x^2+y^2+z^2)^{1/2})
                \theta(R_A-(x^2+(b-y)^2)^{1/2}),
  \end{align}
where
  \begin{equation}
  \theta(x)=\left\{
  \begin{array}{ll}
  0 \hspace{1cm}if \hspace{0.2cm} x<0 \\
  1 \hspace{1cm}otherwise
  \end{array}
  \right.
  \end{equation}
$R_A$ and $\rho_A$ ($R_B$ and $\rho_B$) are the radius and nuclear density
of the nucleus $A$ ($B$), respectively. The nuclear density of nucleus $A$ is
normalized to the atomic number $A$ (we denote the nucleus and its atomic
number by $A$ uniquely). The STAR and PHOBOS collaborations have reported
that their analysed Glauber model number of participant nucleons is nearly
reaction energy independent for a given centrality A+B collisions within 
error bars \cite{star,phob2}. Therefore we assume normal nuclear density of
$\rho_A=\rho_B=\rho_0$=0.16 fm$^{-3}$ here.

The geometric number of spectator nucleons is then calculated by
  \begin{equation}
  N_{spec}^A=A-N_{part}^A, \hspace{0.2cm}N_{spec}^B=B-N_{part}^B.
  \end{equation}

  \subsubsection{Nucleon initiation in the position and momentum phase spaces}
The geometric participant nucleons in each of the colliding nuclei are 
distributed randomly inside the geometric overlap zone. And the geometric 
spectator nucleons are distributed according to the Woods-Saxon distribution 
and $4\pi$ uniform distribution and requiring to be outside the geometric 
overlap zone. These distrbutions are
  \begin{align}
  &\rho(r)=\rho_0(1+\exp{\frac{r-R}{d}})^{-1}, \\
  &f(\theta,\phi)=\frac{1}{4\pi},
  \end{align}
where $R=r_0A^{1/3}$+0.54 fm is the radius of nucleus $A$, the stander 
root-mean-square-radius parameter $r_0 \sim$1.15 fm, and $d\sim$ 0.54 fm is 
the length of diffusion tail.

For the case of collider the $x$ and $y$ components of momentum of each 
nucleons in colliding nucleus are assumed to be zero, provided the beam is 
orientated towards the $z$ direction. The beam momentum is given to $p_z$ 
for each nucleons properly. This means the effect of Fermi motion is 
neglected in the relativistic nuclear collisions.
 \subsection{Particle propagation (cascade) and the collision criteria}
The particles (partons in the parton rescattering stage or hadrons
in the hadron rescattering stage) are assumed to travel in the
straight trajectories
 \begin{equation}
 \vec{r}_i=\vec{R}_i+\vec{\beta}_i(t-T_i)
 \end{equation}
along the momentum direction during two consecutive collisions. In the above
equation ($\vec{R}_i$, $T_i$) and $\vec{\beta}_i$ stand for the four-position
and velocity of the $i$-th particle at the moment of last collision,
respectively.

The particles interact with each other by means of total cross
section. In order to perform the collisions in a consistent way one
needs to introduce a time ordering procedure. We calculate the time
between the last and the next possible collisions for all particle
pairs and then choose the minimum collision time pair to perform the
particle-particle collision \cite{sa1}. A particle-particle
collision (particle $i$ bombards with  $j$, for instance) is defined
to occur provided the minimum approaching distance ($D$) between
particles $i$ and $j$, in their cms system, satisfies \cite{sa1}
 \begin{equation}
 D\leq\sqrt{\sigma_{tot}/\pi},
 \label{dist}
 \end{equation}
where $\sigma_{tot}$ is the total cross section in fm$^2$. We
calculate $D$ starting from the trajectories of particle $i$ and $j$ in
their cms system (variables in the cms system indicated with
superscript $*$)
 \begin{equation}
 \vec{r}_i^*=\vec{R}_i^*+\vec{\beta}_i^*(t^*-T_i^*) \hspace{0.3cm},
        \hspace{0.3cm} \vec{r}_j^*=\vec{R}_j^*+\vec{\beta}_j^*(t^*-T_j^*).
 \end{equation}
The squared relative distance between $i$ and $j$ reads
 \begin{equation}
 d^2(t^*)=(\Delta\vec{R}^*)^2+2\Delta\vec{R}^*\Delta\vec{\beta}^*
 [t^*-\frac{1}{\Delta\vec{\beta}^*}(\vec{\beta}_j^*T_j^*-\vec{\beta}_i^*T_i^*)]
 +(\Delta\vec{\beta}^*)^2[t^*-\frac{1}{\Delta\vec{\beta}^*}(\vec{\beta}_j^*T_j
 ^*-\vec{\beta}_i^*T_i^*)]^2,
 \label{d2t}
 \end{equation}
where
 \begin{equation}
 \Delta\vec{R}^*=\vec{R}_j^*-\vec{R}_i^*,\hspace{0.3cm}
 \Delta\vec{\beta}^*=\vec{\beta}_j^*-\vec{\beta}_i^*.
 \end{equation}

From the requirement of
 \begin{equation}
 \frac{\partial(d^2(t^*))}{\partial t^*}=0
 \end{equation}
one obtains the collision time at the moment of $i$ bombarding with $j$
 \begin{equation}
 t_{ij}^*=\frac{-\Delta\vec{R}^*\Delta\vec{\beta}^*+\Delta\vec{\beta}^*
     (\vec{\beta}_j^*T_j^*-\vec{\beta}_i^*T_i^*)}{(\Delta\vec{\beta}^*
     )^2}.
 \label{tij}
 \end{equation}
Inserting Eq. (\ref{tij}) into Eq. (\ref{d2t}), the corresponding minimum 
approaching distance is obtained
 \begin{equation}
 D=[(\Delta\vec{R}^*)^2+\Delta\vec{R}^*\Delta\vec{\beta}^*(t_{ij}^*-
   \frac{1}{\Delta\vec{\beta}^*}(\vec{\beta}_j^*T_j^*-\vec{\beta}_i^*T_i^*)
   )]^{1/2}.
 \end{equation}

As the particle collision must be causal, the above calculated collision time
should satisfy
 \begin{equation}
 t_{ij}^*\geq\max (T_i^*,T_j^*).
 \label{tijc}
 \end{equation}
Eqs. (\ref{dist}) and (\ref{tijc}) are the criteria for a
particle-particle collision to occur.

This collision time has to boost back from the cms of colliding pair to the
cms of nucleus-nucleus collision according to the inverse Lorentz
transformation \cite{hage,pdg}
  \begin{equation}\label{invlor}
  \begin{split}
  \vec{r}=&\vec{r^{*}}+\vec\beta[\frac{(\gamma-1)\vec{r^{*}}\cdot\vec\beta}
          {\beta^2}+\gamma t^{*}],\\
  t=&\gamma(t^{*}+\vec{r^{*}}\cdot\vec\beta),\\
  \vec\beta=&\frac{\vec{p}_i+\vec{p}_j}{E_i+E_j},\\
  \gamma=&\frac{1}{\sqrt{1-\beta^2}},
\end{split}
\end{equation}
where $\vec\beta$ refers to the velocity of cms of colliding pair in
the frame of the cms of nucleus-nucleus collision. Ordering the
collision times calculated for all collision pairs, the collision
(time) list is obtained.

The corresponding Lorentz transformation reads \cite{hage,pdg}
  \begin{equation*}
  \vec{r^{*}}=\vec{r}+\vec\beta[\frac{(\gamma-1)\vec{r}\cdot\vec\beta}
              {\beta^2}-\gamma t],
  \end{equation*}
  \begin{equation}
  t^{*}=\gamma(t-\vec{r}\cdot\vec\beta).
  \label{lor}
  \end{equation}
Replacing four position ($\vec r$, $t$) with four momentum ($\vec
p$, $E$) in Eq. (\ref{lor}) and (\ref{invlor}) one obtains,
respectively, the Lorentz transformation and inverse Lorentz
transformation for the four momentum.

 \subsection{Cross sections}
  \subsubsection{Hadron-hadron cross sections}
The isospin averaged parametrization formulas are employed for the
hh cross sections \cite{koch,bald}. Taking the $\pi+p\rightarrow k+Y$
($Y$ refers to $\Lambda$ and $\Sigma$) reactions as example, the cross 
sections are

$\pi^+p\rightarrow\Sigma^+k^+$:
\begin{equation}
\sigma(\sqrt{s})=\left\{ \begin{aligned}
         A_1(\sqrt{s}-1.683), \quad &1.683~\rm{GeV}<\sqrt{s}<1.934~\rm{GeV}, \\
         &A_1=0.7~\rm{mb}/0.218~\rm{GeV}, \\
         A_2\exp(-A_3\sqrt{s}),\quad &1.934~\rm{GeV}<\sqrt{s}<3.0~\rm{GeV},\\
         &A_2=60.26~\rm{mb},\quad A_3=2.31~\rm{GeV^{-1}}, \\
         A_4\exp(-A_5\sqrt{s}),\quad &3.0~\rm{GeV}<\sqrt{s},\\
         &A_4=0.36~\rm{mb},\quad A_5=0.605~\rm{GeV^{-1}},  
         \end{aligned} \right.
\end{equation}

$\pi^-p\rightarrow \Lambda k^0$:
\begin{equation}
\sigma(\sqrt{s})=\left\{ \begin{aligned}
         A_1(\sqrt{s}-1.613), \quad &1.613~\rm{GeV}<\sqrt{s}<1.684~\rm{GeV}, \\
         &A_1=0.9~\rm{mb}/0.091~\rm{GeV}, \\
         A_2\exp(-A_3\sqrt{s}),\quad &1.684~\rm{GeV}<\sqrt{s}<2.1~\rm{GeV},\\
         &A_2=436.3~\rm{mb},\quad A_3=4.154~\rm{GeV^{-1}}, \\
         A_4\exp(-A_5\sqrt{s}),\quad &2.1~\rm{GeV}<\sqrt{s},\\
         &A_4=0.314~\rm{mb},\quad A_5=0.301~\rm{GeV^{-1}},
         \end{aligned} \right.
\end{equation}

$\pi^-p\rightarrow \Sigma^0 k^0$:
\begin{equation}
\sigma(\sqrt{s})=\left\{ \begin{aligned}
         A_1(\sqrt{s}-1.689), \quad &1.689~\rm{GeV}<\sqrt{s}<1.722~\rm{GeV}, \\
         &A_1=10.6~\rm{mb}/~\rm{GeV}, \\
         A_2\exp(-A_3\sqrt{s}),\quad &1.722~\rm{GeV}<\sqrt{s}<3.0~\rm{GeV},\\
         &A_2=13.7~\rm{mb},\quad A_3=1.92~\rm{GeV^{-1}}, \\
         A_4\exp(-A_5\sqrt{s}),\quad &3.0~\rm{GeV}<\sqrt{s},\\
         &A_4=0.188~\rm{mb},\quad A_5=0.611~\rm{GeV^{-1}},
         \end{aligned} \right.
\end{equation}

$\pi^-p\rightarrow \Sigma^- k^+$:
\begin{equation}
\sigma(\sqrt{s})=\left\{ \begin{aligned}
         A_6(constant), \quad &1.691~\rm{GeV}<\sqrt{s}<1.9~\rm{GeV}, \\
         A_4\exp(-A_5\sqrt{s}),\quad &1.9~\rm{GeV}<\sqrt{s},\\
         &A_4=309.06~\rm{mb},\quad A_5=3.77~\rm{GeV^{-1}}.
         \end{aligned} \right.
\end{equation}

An assumed constant total cross sections of $\sigma_{\rm{tot}}^{NN}=40$~mb 
(RHIC energy region) or 76~mb (LHC energy region) \cite{pdg}, $\sigma_{\rm
{tot} }^{\pi N}=25$~mb, $\sigma_{\rm{tot}}^{kN}= 21$~mb, and $\sigma_{\rm
{tot}}^{\pi \pi}=10$~mb as well as the inelastic to total cross section ratio 
of 0.85 are provided as another option. We assume
   \begin{equation}
   \sigma_{tot}^{pp}=\sigma_{tot}^{nn}=\sigma_{tot}^{pn}=\sigma_{tot}^{\Delta
   n}=\sigma_{tot}^{\Delta\Delta}.
   \end{equation}
The above hh cross sections are used in the hadron rescattering stage and in 
the parton initiation stage for the nucleus-nucleus collisions as well.
  \subsubsection{Parton-parton cross sections}
We consider only the 2$\to$ 2 parton-parton (re)scattering at the
moment. The nine parton-parton differential cross sections
calculated by means of LO-pQCD \cite{comb,ffb} are:
  \begin{eqnarray}
  \frac{d\hat\sigma}{d\hat t}(ab\rightarrow cd;\hat s,\hat t)=
  \frac{\pi\alpha_s^2}{\hat s^2}|\overline{\sc M}(ab\rightarrow cd)|^2, \\
  |\overline{\sc M}(q_iq_j\rightarrow q_iq_j)|^2=
   \frac{4}{9}\frac{\hat s^2+\hat u^2}{\hat t^2}\approx
   \frac{8}{9}\frac{\hat s^2}{\hat t^2}, \label{proc1} \\
  |\overline{\sc M}(q_iq_i\rightarrow q_iq_i)|^2=
   \frac{4}{9}(\frac{\hat s^2+\hat u^2}{\hat t^2}+\frac{\hat s^2+\hat t^2}
   {\hat u^2})-\frac{8}{27}\frac{\hat s^2}{\hat u\hat t}\approx
   \frac{8}{9}\frac{\hat s^2}{\hat t^2}, \label{proc2} \\
  |\overline{\sc M}(q_i\bar{q}_j\rightarrow q_i\bar{q}_j)|^2=
   \frac{4}{9}\frac{\hat s^2+\hat u^2}{\hat t^2}\approx
   \frac{8}{9}\frac{\hat s^2}{\hat t^2}, \label{proc3} \\
  |\overline{\sc M}(q_i\bar{q}_i\rightarrow q_j\bar{q}_j)|^2=
   \frac{4}{9}\frac{\hat t^2+\hat u^2}{\hat s^2}\approx 0, \label{proc4} \\
  |\overline{\sc M}(q_i\bar{q}_i\rightarrow q_i\bar{q}_i)|^2=
   \frac{4}{9}(\frac{\hat s^2+\hat u^2}{\hat t^2}+\frac{\hat t^2+\hat u^2}
   {\hat s^2})-\frac{8}{27}\frac{\hat u^2}{\hat s\hat t}\approx
   \frac{8}{9}\frac{\hat s^2}{\hat t^2}, \label{proc5} \\
  |\overline{\sc M}(q_i\bar{q}_i\rightarrow gg)|^2=
   \frac{32}{27}\frac{\hat u^2+\hat t^2}{\hat u\hat t}-
   \frac{8}{3}\frac{\hat u^2+\hat t^2}{\hat s^2}\approx
   -\frac{32}{27}\frac{\hat s}{\hat t}, \label{proc6} \\
  |\overline{\sc M}(gg\rightarrow q_i\bar{q}_i)|^2=
   \frac{1}{6}\frac{\hat u^2+\hat t^2}{\hat u\hat t}-\frac{3}{8}
   \frac{\hat u^2+\hat t^2}{\hat s^2}\approx
   -\frac{1}{3}\frac{\hat s}{\hat t}, \label{proc7} \\
  |\overline{\sc M}(q_ig\rightarrow q_ig)|^2=
   -\frac{4}{9}\frac{\hat u^2+\hat s^2}{\hat u\hat s}+\frac{\hat u^2+\hat s^2}
   {\hat t^2}\approx \frac{2\hat s^2}{\hat t^2}, \label{proc8} \\
  |\overline{\sc M}(gg\rightarrow gg)|^2=
   \frac{9}{2}(3-\frac{\hat u\hat t}{\hat s^2}-\frac{\hat u\hat s}{\hat t^2}
   -\frac{\hat s\hat t}{\hat u^2})\approx\frac{9}{2}\frac{\hat s^2}{\hat t^2}
   \label{proc9}.
  \end{eqnarray}
Here the parton is represented as a classical point-like particle
and specified by its internal degree of freedoms: flavor, rest
(current) mass, momentum, and position. The spin and color degrees
of freedoms are not explicitly included \cite{mull}. Therefore the
amplitudes $|\overline{\sc M}|^2$ are calculated by averaging over
spin and color degrees of freedoms. The parton energy is determined
by $E^2=p^2+m^2+M^2$, where $M$ refers to virtual mass
($M^2$=0 for on mass shell parton, $M^2<$0 for space-like parton,
and $M^2>$0 for time-like parton) \cite{mull}. $\alpha_s$ refers to
the effective strong coupling constant. $\hat s$, $\hat t$, and
$\hat u$ stand for the Mandelstam variables. In each of the above
equations the second form is obtained by keeping only the leading
divergent terms \cite{bin} with the assumption of the parton current
mass is negligible relative to the cms energy. This assumption leads to
  \begin{equation}
  \hat s+\hat t+\hat u=0.
  \end{equation}
The subprocesses of Eqs. (\ref{proc4}), (\ref{proc6}), and (\ref{proc7}) are
inelastic (re)scatterings, otherwise elastic. Subprocess of Eq. (\ref{proc4})
is always negligible.

If the thermal direct photon is interested following two subprocesses
have to be included:
  \begin{eqnarray}
  |\overline{\sc M}(q\bar{q}\rightarrow g\gamma)|^2=
   \frac{8}{9}\frac{\alpha}{\alpha_s}e_q^2(\frac{\hat u}{\hat t}+
   \frac{\hat t}{\hat u}) \approx
   -\frac{16}{9}\frac{\alpha}{\alpha_s}e_q^2\frac{\hat s}{\hat t}
   \label{proc10}, \\
  |\overline{\sc M}(qg\rightarrow q\gamma)|^2=-\frac{1}{3}\frac{\alpha}
   {\alpha_s}e_q^2(\frac{\hat t}{\hat s}+\frac{\hat s}{\hat t}) \approx
   -\frac{1}{3}\frac{\alpha}{\alpha_s}e_q^2\frac{\hat s}{\hat t}
  \label{proc11},
  \end{eqnarray}
where $\alpha$=1/137 stands for the fine-structure constant, $e_q$ refers to
the charge of quark $q$.
 \subsection{Determination of the scattered particles}
The particle-particle (re)scattering is always dealt in the cms of
colliding pair. The superscript $*$ is neglected in the rest of this
section, except mentioned specially.
  \subsubsection{Hadron-hadron collisions}
For an inelastic hadron (re)scattering $1+2\rightarrow 3+4~(m_1\neq
m_3, m_2\neq m_4)$, the four momentum of the scattered particle
reads
  \begin{align}\label{fourm}
  |\vec{p}_3|^2=&|\vec{p}_4|^2=\frac{[s-(m_3+m_4)^2][s-(m_3-m_4)^2]}{4s},
  \nonumber\\
  \epsilon_3=&\frac{s+(m_3^2-m_4^2)}{2\sqrt{s}},  \\
  \epsilon_4=&\frac{s-(m_3^2-m_4^2)}{2\sqrt{s}}.\nonumber
  \end{align}
They are deduced from four momentum conservation \cite{hage,pdg}. In the 
above equation $s$ refers to the cms energy squared.

The momentum orientation (polar angle $\theta$ and azimuthal angle
$\phi$) of scattered particle is decided by the assumption that the
momentum transfer squared, $t$, is distributed as \cite{cung,sa1}
  \begin{eqnarray}
  \frac{d\sigma}{dt}\sim\exp{Bt}, \label{tt11}\\
  \alpha=3.65(\sqrt{s}-m_1-m_2)\approx 3.65\sqrt{s} , \label{tt13}\\
  B=\frac{\alpha^6}{1+\alpha^6}A\approx A , \label{tt12}\\
  A=\min(10.3,(1.12<p_T>)^{-2}),   \label{tt14}
  \end{eqnarray}
where $<p_T>$ stands for the mean transverse momentum assumed to be
0.5 GeV/c because it is observed in the experiment that the value of
$<p_T>$ may be not so sensitive to the reaction energy and
centrality \cite{alice}. In the above equations the second form was
obtained provided the particle mass is negligible for high cms
energy. On the other hand, $t$ is related to $\theta_s$ spanned between 
$\vec{p}_1$ and $\vec{p}_3$ \cite{pdg} by
  \begin{align}
  &t=m_1^2-2\epsilon_1\epsilon_3+2\vec{p}_1\cdot\vec{p}_3+m_3^2,
   \label{tt21}\\
  &t_{max}=m_1^2-2\epsilon_1\epsilon_3+2|\vec{p}_1||\vec{p}_3|+m_3^2,
   \label{tt22}\\
  &t_{min}=m_1^2-2\epsilon_1\epsilon_3-2|\vec{p}_1||\vec{p}_3|+m_3^2,
   \label{tt23}
  \end{align}
where $t_{max}$ and $t_{min}$ are corresponding to $\cos{\theta_s}$=1 and -1, 
respectively. Therefore we first sample a $t'$ value within [$t_{min}$, 
$t_{max}$] according to exponential distribution Eq. (\ref{tt11})
  \begin{equation}
  t'=\frac{1}{B}\ln[\xi\exp(Bt_{max})+(1-\xi)\exp(Bt_{min})].
  \label{tt31}
  \end{equation}
The corresponding $\cos{\theta_s}$ is then calculated by Eq. (\ref{tt21})
  \begin{equation}
  \cos{\theta_s}=\frac{0.5(t'-m_1^2-m_3^2)+\epsilon_1\epsilon_3}
                  {|\vec{p}_1||\vec{p}_3|}.
  \label{tt32}
  \end{equation}
The azimuthal angle $\phi_s$ is sampled  isotropically in 2$\pi$.

It should be mentioned that the orientation of scattered particle 3,
for instance, is relative to scattering particle 1, so it should
rotate back to the particle 1 and 2 cms system, where particle 1
is described by $|\vec{p}_1|$, $\theta_1$, and $\phi_1$. This
rotation reads
  \begin{align}\label{rot}
  p^{'}_{3x}=&p_{3x}[\cos{\phi_1}(\cos{\theta_1}\sin{\theta_s}\cos{\phi_s}
           +\sin{\theta_1}\cos{\theta_s})-\sin{\phi_1}\sin{\theta_s}
           \sin{\phi_s}], \nonumber\\
  p^{'}_{3y}=&p_{3y}[\sin{\phi_1}(\cos{\theta_1}\sin{\theta_s}\cos{\phi_s}
           +\sin{\theta_1}\cos{\theta_s})+\cos{\phi_1}\sin{\theta_s}
           \sin{\phi_s}], \\
  p^{'}_{3z}=&p_{3z}[\cos{\theta_1}\cos{\theta_s}-
           \sin{\theta_1}\sin{\theta_s}\cos{\phi_s}]\nonumber.
  \end{align}
At last the scattered particles are boosted back to the moving frame of 
nucleus-nucleus cms system.

It is impossible to take all inelastic channels into account, only
following inelastic reactions
  \begin{eqnarray*}
\pi N\rightleftharpoons\pi\Delta\hspace{1cm}\pi N\rightleftharpoons\rho N,\\
NN\rightleftharpoons N\Delta\hspace{1cm}\pi\pi\rightleftharpoons k\bar k, \\
\pi N\rightleftharpoons kY\hspace{1cm}\pi\bar N\rightleftharpoons\bar k\bar Y,
 \\
\pi Y\rightleftharpoons k\Xi\hspace{1cm}\pi\bar Y\rightleftharpoons\bar k
  \bar\Xi, \\
\bar k N\rightleftharpoons\pi Y\hspace{1cm}k\bar N\rightleftharpoons\pi\bar Y
 ,\\
\bar kY\rightleftharpoons\pi\Xi\hspace{1cm}k\bar Y\rightleftharpoons\pi\bar
 \Xi, \\
\bar kN\rightleftharpoons k\Xi\hspace{1cm}k\bar N\rightleftharpoons\bar k\bar
  \Xi, \\
\pi\Xi\rightleftharpoons k\Omega^-\hspace{1cm}\pi\bar\Xi\rightleftharpoons\bar
  k\bar\Omega^-,\\
k\bar\Xi\rightleftharpoons\pi{\bar\Omega}^-\hspace{1cm}\bar k\Xi
  \rightleftharpoons\pi\Omega^-,\\
N\bar N annihilation,
N\bar Y annihilation,
  \end{eqnarray*}
are considered. The rest is treated as elastic reactions in a sense. Even so, 
there are already $\sim$600 inelastic channels involved.

Taking $\pi N$ (re)scattering as an example, there are channels of
  \begin{equation}
  \pi N\rightarrow\pi\Delta, \hspace{0.4cm}
  \pi N\rightarrow\rho N, \hspace{0.4cm}
  \pi N\rightarrow kY,
  \end{equation}
therefore the relative probabilities of these channels are invoked
to select one among them. The cross section of the reverse reactions
are calculated by means of detailed balance.

The final state of $N\bar N$ and $N\bar Y$ annihilations are simply
treated as five particle state through $N\bar N\rightarrow\rho\omega
\rightarrow 5\pi$ and $N\bar Y\rightarrow k^*\omega\rightarrow
k+4\pi$, respectively. The cross section of $N\bar Y$ annihilation
is assumed to be 1/5 of the $N\bar N$ annihilation.

For elastic hadron (re)scattering because
  \begin{equation}
  m_1=m_3, \hspace{0.4cm} m_2=m_4, \hspace{0.4cm}
  |\vec{p}_1|=|\vec{p}_2|=|\vec{p}_3|=|\vec{p}_4|, \hspace{0.4cm}
  \epsilon_1=\epsilon_3, \hspace{0.4cm} \epsilon_2=\epsilon_4,
  \end{equation}
Eqs. (\ref{tt21})-(\ref{tt32}) reduce, respectively, to
  \begin{align}
  &t=2|\vec{p}_1|^2(\cos{\theta_s}-1),\label{tt41}\\
  &t_{max}=0, \label{tt42} \\
  &t_{min}=-4|\vec{p}_1|^2, \label{tt43}
  \end{align}
  \begin{equation}
  t'=\frac{1}{B}\ln[\xi+(1-\xi)\exp(Bt_{min})],  \label{tt51}
  \end{equation}
  \begin{equation}
  \cos{\theta_s}=1+\frac{t'}{2|\vec{p}_1|^2}. \label{tt52}
  \end{equation}
The rest is the same as that of inelastic (re)scattering.
  \subsubsection{Parton-parton collisions}
If several partonic final states can be reached by a single
partonic initial state the following relative probability is invoked to
select one among them
  \begin{equation}
  \hat{\sigma}(ab\rightarrow cd;\hat s)/\hat{\sigma}_{ab}(\hat s)
  \end{equation}
where
  \begin{align}
  \hat{\sigma}(ab\rightarrow cd;\hat s)=&
  \int_{-\hat s}^0\frac{d\hat\sigma}{d\hat t}(ab\rightarrow cd;\hat s,\hat t)
   d\hat t \\
  \hat{\sigma}_{ab}(\hat s)=&\sum_{c,d}\hat{\sigma}(ab\rightarrow cd;\hat s)
   \label{ptot}.
  \end{align}

In the inelastic parton-parton (re)scattering the flavor of scattered quark
(antiquark) is decided according to
  \begin{equation}
  u:d:s:c\cdots=\gamma_u:\gamma_d:\gamma_s:\gamma_c\cdots
  \label{gama}
  \end{equation}
where $\gamma_u,\gamma_d,\gamma_s,\gamma_c,\cdots$ are input
parameters and their default values are 1:1:0.3:$10^{-11}\cdots$.
The $c$ quark is not included presently.

The momentum transfer squared $\hat t$ is sampled according to
  \begin{equation}
  \Phi(\hat t)=\frac{1}{\hat{\sigma}(ab\rightarrow cd;\hat s)}
   \int_{-\hat s}^{\hat t}\frac{d\hat\sigma}{d\hat t^\prime}(ab\rightarrow
   cd;\hat s,\hat t^\prime)d\hat t^\prime.
  \end{equation}
Then the (re)scattering angle is calculated by \cite{hage,pdg}
  \begin{equation}
  \cos\theta_s=1+\frac{2\hat s\hat t}{[\hat s-(m_a+m_b)][\hat s-(m_a-m_b)]}
   \approx 1+\frac{2\hat t}{\hat s}=1+\frac{\hat t}{2|\vec{p}_a|^2},
  \end{equation}
where the second form is obtained provided the parton mass is
negligible relative to the cms energy. The azimuthal angle is
sampled randomly in $2\pi$.
 \subsection{Diquark break-up}
In order to obtain the initial partonic state for nuclear collision system 
one has to switch-off string fragmentation in PYTHIA, to break-up the 
strings, and to break-up the diquarks in the strings. The diquark break-up 
is performed in its rest frame according to two-body decay kinematics 
\cite{hage,pdg}. The energy and momentum modulus of broken quarks are 
calculated by Eq. (\ref{fourm}) with $m_3$ and $m_4$ identified as the mass 
of broken quarks and $s$ the squared four momentum of diquark. The 
orientation of one of the broken quarks is sampled isotropically in 4$\pi$ 
and the other is just orientated oppositely.

In the above, the three momentum of broken quark is calculated
relative to three momentum of diquark, so it must be rotated back to
the frame where diquark is described according to Eq. (\ref{rot}).
It is also need to boost back to the moving frame of diquark
according to inverse Lorentz transformation Eq. (\ref{invlor}).

One of the broken quarks is assumed to have the diquark four
position. The other is assumed to be created at the same time but is
arranged randomly around the first one within 0.5 fm in each of the three 
position coordinates.

The gluons are also needed to be broken-up in the hadronization by
the Monte Carlo coalescence model. Because of zero mass the gluon is
broken according to three momentum conservation. The energy
discrepancy between the breaking gluon and the broken quarks is
shared among quarks and antiquarks. The flavor of broken quark is decided by 
Eq. (\ref{gama}).
 \subsection{Reduction of the strange (heavy) quark suppression}
In the LUND string fragmentation regime the $q\bar q$ pair with quark mass
$m$ and transverse momentum $p_T$ may be created quantum mechanically at one
point and then tunnel out to the classically allowed region \cite{sjo1}. This 
tunnelling probability is given by
 \begin{equation}
 \exp(-\frac{\pi m^2}{\kappa})\exp(-\frac{\pi p_T^2}{\kappa})
 \label{tunn}
 \end{equation}
where the string tension $\kappa$ is assumed to be $\approx$1 
GeV/fm $\approx$0.2 GeV$^2$ \cite{sjo1}. This probability implies a 
suppression of strange (heavy) quark production of $u:d:s:c \approx 1:1:0.3:
10^{-11}$. The charm and heavier quarks are not expected to be produced in 
the soft string fragmentation process if there is no charm and heavier quarks 
in the string originally. They are expected to be produced only in the hard 
process or as a part of the initial- and final-state QCD radiations. The 
higher $p_T$ quark pair is expected to be created provided the string tension 
is large. However, in the PYTHIA 6.4 model there are model parameters of
 \begin{itemize}
 \item parj(1) is the suppression of diquark-antidiquark pair production
compared with quark-antiquark production,
 \item parj(2) is the suppression of $s$ quark pair production compared with
$u$ or $d$ pair production,
 \item parj(3) is the extra suppression of strange diquark production
compared with the normal suppression of strange quark,
 \item parj(21) corresponds to the width $\sigma$ in the Gaussian $p_x$ and
$p_y$ transverse momentum distributions for primary hadrons.
 \end{itemize}
They are able to be tuned to reduce the strange quark suppression and to 
change the width of its $p_T$ distribution.

We introduced a mechanism of the increase of effective string tension and
hence the reduction of strange quark suppression in \cite{sa3}. In
that paper we assumed that the effective string tension increases with the
increasing of the number and hardening of gluons in the string by
 \begin{equation}
 \kappa^{eff}=\kappa_0(1-\xi)^{-a},
 \label{kap}
 \end{equation}
 \begin{equation}
 \xi=\frac{\ln(\frac{k_{Tmax}^2}{s_0})}{\ln(\frac{s}{s_0})+\sum_{j=2}^{n-1}
     \ln(\frac{k_{Tj}^2}{s_0})}.
 \label{xii}
 \end{equation}
In above equations, $\kappa_0$ is the string tension of the pure $q\bar q$ 
string assumed to be $\sim$ 1 GeV/fm. The gluons in the multigluon string are 
ordered from 2 to n-1 because 1 and n refer to quark and antiquark at two 
ends of string. $k_{Tj}$ is the transverse momentum of gluon $j$ with $k_{Tj}
^2\geq s_0$ and $k_{Tmax}$ is the largest transverse momentum among the 
gluons. The parameters $a$=3.5 GeV and $s_0$=0.8 GeV were determined by 
fitting the hh collision data. It should be mentioned that Eq. (\ref{xii}) 
represents the deviation scale of the multigluon string from the pure string.

If $\lambda$ denotes parj(2) (parj(1), parj(3)) then by Eq. (\ref{tunn}) the 
following relations between two strings with, respectively, effective string 
tension of $\kappa^{eff}_1$ and $\kappa^{eff}_2$ are obtained
 \begin{equation}
 \lambda_2=\lambda_1^{\frac{\kappa^{eff}_1}{\kappa^{eff}_2}},
 \label{lamd}
 \end{equation}
 \begin{equation}
 \sigma_2=\sigma_1(\frac{\kappa^{eff}_2}{\kappa^{eff}_1})^{1/2},
 \end{equation}
where subscripts identify the strings.

Above mechanism is involved in the PACIAE 2.0 model. Therefore one can first 
tune the parameters of parj(1), (2), (3), and (21) to fit the strangeness 
production data in a given nuclear collision at a given energy. Then the 
resulted parj(1), (2), (3), (21) and effective string tension can be used to 
predict the strangeness production in the same reaction system at different 
energy even in the different reaction systems.
 \subsection{Deexcitation of the energetic quark (antiquark) in coalescence
             model}
If the energy of a quark (anti-quark) is large enough (e.g. $>$2
GeV), one should introduce deexcitation mechanism for this quark
(antiquark) in the coalescence model. This is similar to the
iterative approach of quark-antiquark pair generation in the LUND
string fragmentation regime \cite{sjo,sjo1,sjo2}. Taking an
energetic initial quark $q_0$ as an example, a new $q_1\bar{q}_1$
pair may be created from vacuum. This quark-antiquark pair brings a
fraction $z$ of energy (momentum) of $q_0$ and leaves remnant of
$q_0$ behind. If the energy of $q_0$ remnant is still large enough,
a new $q_2\bar{q}_2$ pair creates and above processes repeats until
the energy is not enough to generate a new quark-antiquark pair.

In the case of positive longitudinal momentum of mother quark
$p_{0z}$ in order to obtain four momentum of the first daughter pair
$q_1\bar{q}_1$ we introduce the forward light cone variable
  \begin{equation}
  W_0=E_0+p_{0z}.
  \label{a66}
  \end{equation}
The LUND string fragmentation function
  \begin{equation}
  f(z)\propto z^{-1}(1-z)^{\alpha}\exp({-\beta m_T^2/z})
  \end{equation}
with default values of $\alpha=0.3$ and $\beta=0.58$ or the Field-Feynman
parametrization
  \begin{equation}
  f(z)\propto 1-\alpha+3\alpha(1-z)^2
  \end{equation}
with default value of $\alpha=0.77$ is used to sample randomly a
fraction variable $z_1$ for the first daughter pair $q_1\bar{q}_1$.
Its forward light cone variable is then
  \begin{equation}
  W_1=z_1W_0.
  \label{a69}
  \end{equation}
From Eqs.~(\ref{a66}) and (\ref{a69}), we obtain
  \begin{eqnarray}
  E_1=\frac{1}{2}(W_1+\frac{m_{1T}^2}{W_1}), \\
  p_{1z}=\frac{1}{2}(W_1-\frac{m_{1T}^2}{W_1}).
  \end{eqnarray}

Similarly, for the negative longitudinal momentum of mother quark, one has to
introduce the backward light cone variable
  \begin{equation}
  w_0=E_0-p_{0z}
  \end{equation}
and has the solution of
  \begin{eqnarray}
  E_1=\frac{1}{2}(W_1+\frac{m_{1T}^2}{W_1}), \\
  p_{1z}=-\frac{1}{2}(W_1-\frac{m_{1T}^2}{W_1}),
  \end{eqnarray}
for the first daughter pair $q_1\bar{q}_1$.

The $p_{1x}$ and $p_{1y}$ of $q_1\bar q_1$ pair are sampled
according to the two dimensional Gaussian distribution \cite{pi}
  \begin{equation}
  \exp[-p_T^2/\sigma]dp_T^2, \hspace{3cm} 0<p_T^2<p_{Tmax},
  \end{equation}
with default values of $\sigma$=0.5 (GeV/c)$^{-2}$, $p_{Tmax}$=6
GeV/c. Then we obtain
  \begin{equation}
  p_{1x}=p_{1T}\cos(\phi), \hspace{3cm} p_{1y}=p_{1T}\sin(\phi),
  \end{equation}
with $\phi$ sampled isotropically in 2$\pi$.

The $q_1\bar q_1$ pair is then broken up in its rest frame according
to two-body kinematics mentioned above. The flavor of quark
(antiquark) is sampled according to Eq. (\ref{gama}). It is also
needed first to rotate back to the frame where $q_1\bar q_1$ pair is
described according to Eq. (\ref{rot}) and then to boost back to the
moving frame of $q_1\bar q_1$ pair according to inverse Lorentz
transformation Eq. (\ref{invlor}).
\section{Program (model) structure}

PACIAE 2.0 has three versions: PACIAE 2.0a (its program packets are 
compressed to 20a.tar.gz) describing the relativistic elementary collisions
(pp, $\rm{\bar pp}$, or e$^+$e$^-$) and PACIAE 2.0b (20b.tar.gz) as well as 
PACIAE 2.0c (20c.tar.gz) describing the relativistic nucleus-nucleus (A+B,
including p+A) collisions.

 \subsection {PACIAE 2.0a structure}

The PACIAE 2.0a model is for elementary collisions, it differs from 
PYTHIA 6.4 in the addition of the parton rescattering before hadronization
and the hadron rescattering after hadronization. In order to create the 
initial partonic state we switch off the string fragmentation temporarily in 
the PACIAE 2.0 model. Then we obtain a partonic initial state (parton list) 
after the strings are broken up and the diquarks (anti-diquarks) are split up 
randomly. This partonic matter proceeds parton rescattering. After parton
rescattering the partonic matter is hadronized by the string fragmentation or 
Monte Carlo coalescence model. The hadronic rescattering is followed. 
Therefore the PACIAE 2.0a, as well as PACIAE 2.0b and PACIAE 2.0c, consists 
of the parton initiation, parton evolution (cascade, rescattering), 
hadronization, and hadron evolution (cascade, rescattering) four stages.

PACIAE 2.0a program consists of paciae\_20a.f, parcas\_20a.f, sfm\_20a.f,
coales\_20a.f, hadcas\_20a.f, and p\_20a.f:
  \begin{enumerate}
  \item paciae\_20a.f plays both functions of an example of user program and
parton initiation. The later is performed by PYTHIA with the
string fragmentation switched-off and the diquarks broken-up
randomly. We obtain a configuration of quarks, anti-quarks, gluons,
and a few hadronic remnants. This is a partonic initial state
(parton list) for an elementary collision.
  \item parcas\_20a.f performs the parton rescattering based on the parton
list above. The key ingredients of this cascade process were:
   \begin{itemize}
   \item The partons are assumed traveling on the straight trajectories along
their momentum direction.
   \item The partons interact with each other by the 2 $\rightarrow$ 2 LO
pQCD cross section $\sigma_{tot}$ \cite{comb,ffb} or its variety of
keeping only leading divergent terms. With the collision times of
all possible interaction pairs the collision time list is composed.
   \item A parton-parton collision pair with least collision time is selected
from the collision time list and executed according to the LO pQCD
differential and total cross sections.
   \item Update the parton list by removing the scattering partons and adding
the scattered (generated) partons.
   \item Update the collision time list by removing the collision pairs
containing any one of the scattering partons and adding the new collision
pairs constructed by one of the scattered (generated) partons with another
one in parton list.
   \item A new parton-parton collision pair with least collision time is
selected from the collision time list.
   \item Repeat last four items until the collision time list is empty 
(partonic freeze-out).
   \end{itemize}
  \item sfm\_20a.f executes the hadronization of partonic matter according
to the LUND string fragmentation after the parton rescattering and
string reconstruction by calling p20a.f.
  \item coales\_20a.f performs the hadronization of partonic matter according
to the Monte Carlo coalescence model after the parton rescattering.
The key ingredients of this coalescence model are:
   \begin{itemize}
   \item The energetic partons are first deexcited.
   \item The gluons are forcibly split into $q\bar q$ pair randomly.
   \item A hadron table, composed of mesons and baryons, inputted to the
program. The pseudoscalar and vector mesons made of u, d, s, and c
quarks, the $B^+$, $B^0$, $B^{*0}$, as well as $\Upsilon$ are
considered as mesons. The baryons include the SU(4) multiplets of
baryons made of u, d, s, and c quarks (except those with double c
quarks) and $\Lambda^0_b$.
   \item Two partons coalesce a meson and three partons a baryon (antibaryon)
in hadron table according to the quark flavor structure of the coalesced
hadron and the flavors, positions, as well as momenta of the coalescing
partons.
   \item If the coalescing partons can form either a pseudoscalar meson or a
vector meson (e. g. $u\bar d$ can be a $\pi^+$ or a $\rho^+$) the
principle of less discrepancy between the invariant mass of
coalescing partons and the mass of coalesced hadron is invoked to
select one from them. The same principle is used in the baryon
production (such as $p$ and $\Delta^+$ are composed of $uud$).
   \item The momentum conservation is required.
   \item The phase space constraint
   \begin{equation}
   \frac{16\pi^2}{9}\Delta r^3\Delta p^3=\frac{h^3}{d}
   \end{equation}
is introduced as an option. In the above equation $h^3/d$ is the volume
occupied by a single hadron in the phase space, $d$=4 refers to the spin and
parity degeneracies of hadron, $\Delta r$ and $\Delta p$ stand for the
position and momentum distances between coalescing partons, respectively.
   \end{itemize}
One selects sfm\_20a.f or coales\_20a.f to hadronize the partonic
matter by  switch parameter adj1(12).
  \item hadcas\_20a.f executes hadron rescattering after hadronization.
This hadron cascade proceeds in the same way as the one in the parton 
cascade mentioned above. But one should note:
   \begin{itemize}
   \item The hadrons of $\pi, k, p, n, \rho (\omega), \Delta, \Lambda, \Sigma,
\Xi, \Omega, J/\Psi$ and their antiparticles are considered here
instead of parton there.
   \item Instead of LO pQCD parton-parton cross section is the
hh total cross section.
   \end{itemize}
  \item p20a.f is different from PYTHIA 6.4 in the addition of  mechanism for 
the reduction of strange quark suppression.
  \item usux.dat is an example for input file. Each program packet (paciae
\_20a.f, parcas\_20a.f, ...) is flexibly to be modified and/or replaced. A 
run can stop at the end of any packet for different purposes. For instance, 
one just sets the switch parameter adj1(40)=1 if one takes interest in the 
partonic output before parton rescattering.
  \end{enumerate}

\tikzset{every picture/.style={very thick},
arrow/.style={-stealth'}
}
\tikzstyle{decision} = [shape aspect=3, diamond, draw, fill=blue!10,
text width=12em, text badly centered, node distance=1.7cm, inner sep=0pt]
\tikzstyle{block} = [rectangle, draw, fill=blue!10,
text width=12em, text centered, rounded corners, minimum height=3em]
\tikzstyle{line} = [draw, -latex']
\tikzstyle{cloud} = [draw, ellipse,fill=red!20, node distance=3cm,
minimum height=2em]

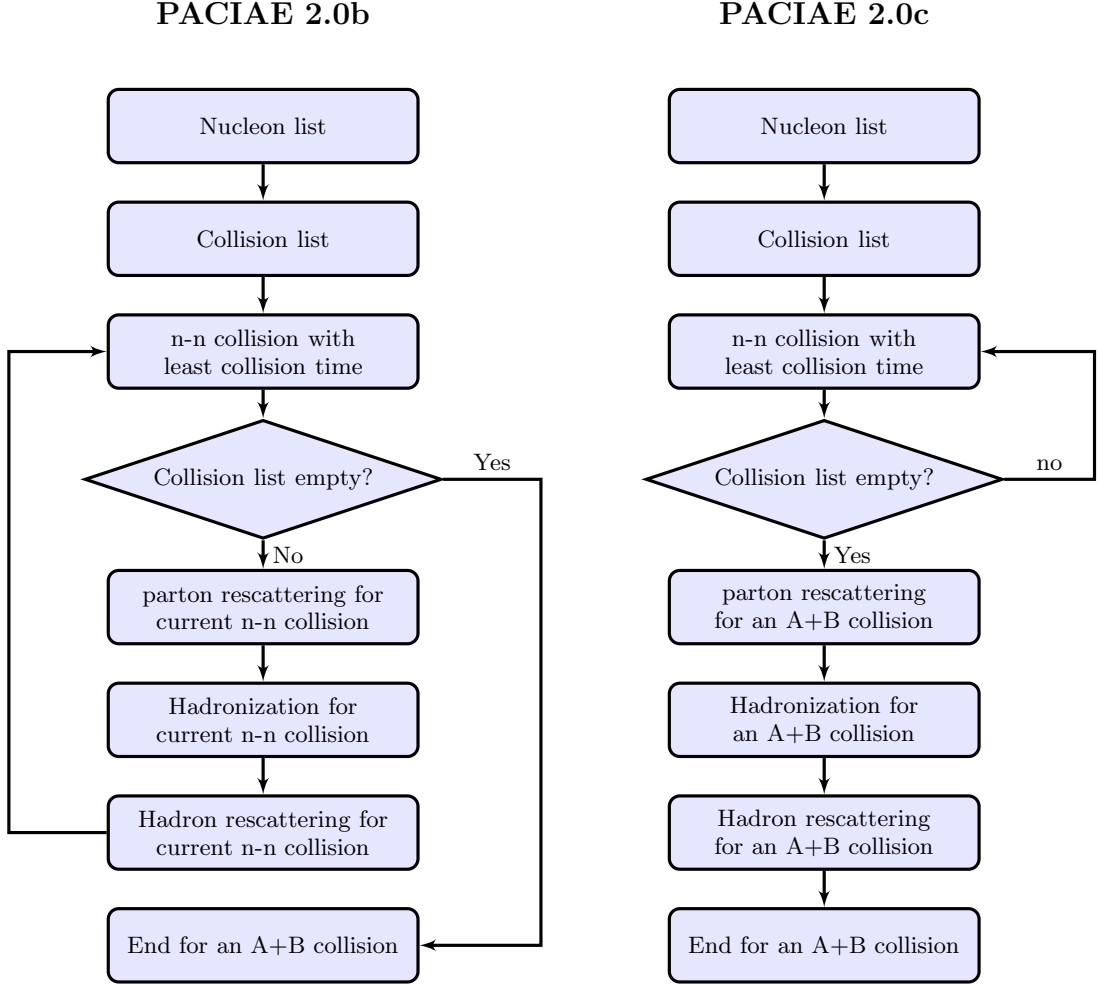
\begin{figure*}
\begin{minipage}[t]{0.45\textwidth}
\centering
\begin{tikzpicture}[node distance = 1.5cm, auto]
\node (PACIAE2b) {\bf \large PACIAE 2.0b};
\node [block, below of=PACIAE2b] (NuList) {Nucleon list};
\node [block, below of=NuList] (CoList) {Collision list};
\node [block, below of=CoList] (nnCol) {n-n collision with least collision
time };
\node [decision, below of=nnCol] (IfCoListEmpty) {Collision list empty?};
\node [block, below of=IfCoListEmpty, node distance = 1.7cm] (PartonRecatter)
{parton rescattering for current n-n collision};
\node [block, below of=PartonRecatter] (Hadronization) {Hadronization for
current n-n collision};
\node [block, below of=Hadronization] (HadronRescattering) {Hadron
rescattering for current n-n collision};
\node [block, below of=HadronRescattering] (End) {End for an A+B collision};
\path [line] (NuList) -- (CoList);
\path [line] (CoList) -- (nnCol);
\path [line] (nnCol) -- (IfCoListEmpty);
\path [line] (IfCoListEmpty) -- node {No} (PartonRecatter) ;
\path [line] (PartonRecatter) -- (Hadronization);
\path [line] (Hadronization) -- (HadronRescattering);
\path [line] (HadronRescattering.west) -- ++(-1.3,0) |- (nnCol.west);
\path [line] (IfCoListEmpty.east) -- node{Yes} ++(1.3,0) |- (End.east);
\end{tikzpicture}
\end{minipage}
\hspace{0.5cm}
\begin{minipage}[t]{0.45\textwidth}
\centering
\begin{tikzpicture}[node distance = 1.5cm, auto]
\node (PACIAE2c) {\bf \large PACIAE 2.0c};
\node [block, below of=PACIAE2c] (NuList) {Nucleon list};
\node [block, below of=NuList] (CoList) {Collision list};
\node [block, below of=CoList] (nnCol) {n-n collision with least collision
time };
\node [decision, below of=nnCol] (IfCoListEmpty) {Collision list empty?};
\node [block, below of=IfCoListEmpty, node distance = 1.7cm] (PartonRecatter)
{parton rescattering for an A+B collision};
\node [block, below of=PartonRecatter] (Hadronization) {Hadronization for an
A+B collision};
\node [block, below of=Hadronization] (HadronRescattering) {Hadron
rescattering for an A+B collision};
\node [block, below of=HadronRescattering] (End) {End for an A+B collision};
\path [line] (NuList) -- (CoList);
\path [line] (CoList) -- (nnCol);
\path [line] (nnCol) -- (IfCoListEmpty);
\path [line] (IfCoListEmpty.east) -- node {no} ++(1.2,0) |- (nnCol.east);
\path [line] (IfCoListEmpty) -- node {Yes} (PartonRecatter) ;
\path [line] (PartonRecatter) -- (Hadronization);
\path [line] (Hadronization) -- (HadronRescattering);
\path [line] (HadronRescattering) -- (End);
\end{tikzpicture}
\end{minipage}
\caption{Structure of PACIAE 2.0b (left) and PACIAE 2.0c (right) for
the dynamical simulation of a nucleus-nucleus (A+B) collision.}
\label{642}
\end{figure*}

\begin{figure}[ht]
\epsfig{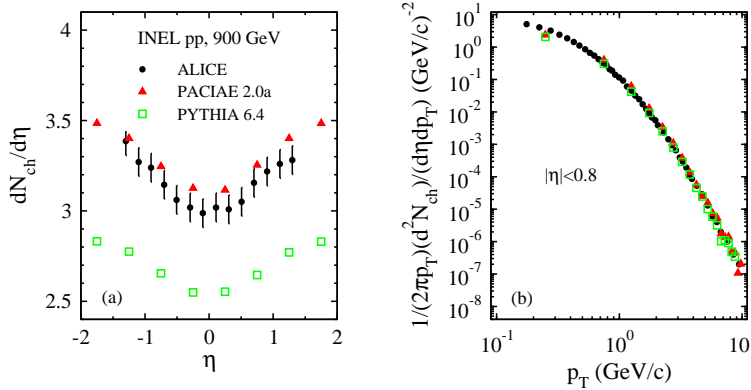} \caption{(Color
online) (a) charged particle pseudorapidity distribution
         in INEL pp collisions at $\sqrt{s}$=900 GeV measured by ALICE
         \cite{alice2} and compared with the PACIAE 2.0a and PYTHIA 6.4
         results. (b) transverse momentum distribution where data taken from
         \cite{alice3}.}
\label{pp900}
\end{figure}

\begin{table}[!htbp]
\centering \caption{Charged particle pseudorapidity densities at mid-rapidity
($|\eta|<1$) for the INEL pp collisions having at least one charged particle
in the same region (INEL$>0_{|\eta|<1}$) at $\sqrt s$=0.9, 2.36, and 7 TeV
measured by ALICE \cite{alice1}. The first data error is statistical and the
second systematic. The corresponding results calculated by PYTHIA 6.4 and
PACIAE 2.0a were given as well.}
\begin{tabular}{cccc}
\hline\hline
$\sqrt{s}$ TeV& ALICE& PYTHIA& PACIAE 2.0a\\
\cmidrule[0.25pt](l{0.05cm}r{0.05cm}){1-1}
\cmidrule[0.25pt](l{0.05cm}r{0.05cm}){2-4}
0.9& 3.81$\pm$0.01$\pm$0.07& 3.35& 3.62\\
2.36 & 4.70$\pm$0.01$\pm$0.10& 4.07 & 4.45\\
7 & 6.01$\pm$0.01$\pm$0.16& 5.25& 5.95\\
\hline\hline
\end{tabular}
\label{mul}
\end{table}
 \subsection {PACIAE 2.0b structure}

PACIAE 2.0b program consists of paciae\_20b.f, parini\_20b.f, parcas\_20b.f, 
sfm\_20b.f, colaes\_20b.f, hadcas\_20b.f, and p20b.f:
  \begin{enumerate}
  \item paciae\_20b.f is a user program.
  \item parini\_20b.f administrates the generation of an event for an A+B
collision:
   \begin{itemize}
   \item Initiation in the position and momentum spaces for an A+B
   collision:
    \begin{itemize}
    \item One colliding nucleus is set on the origin in the position space 
and the other on the position apart from the origin by an impact parameter 
$b$ in the $x$ axis. Both nucleus centers are assumed having $y=z=0$. The 
colliding nucleus is assumed geometrically as a sphere with radius $R_{A(B)}
$ fm.
    \item The geometric number of participant nucleons is first calculated.
Then the participant nucleons are arranged randomly in the overlap region of 
the colliding nuclei. The rest nucleons in the colliding nucleus are randomly 
arranged in the corresponding sphere according to the Woods-Saxon 
distribution (for radius $r$) and 4$\pi$ isotropic distribution (for 
orientation) and are required to be outside the overlap region.
    \item The beam momentum is given to the nucleons in the colliding nucleus
properly.
    \item An initial nucleon list for an A+B collision is then obtained.
    \end{itemize}
\begin{table}[htbp]
\centering \caption{Total charged multiplicity in 0-6\% most central
Au+Au collisions at $\sqrt{s_{NN}}$=0.2 GeV \cite{phob} and the
charged particle pseudorapidity density at mid-rapidity
($|\eta|<$0.5) in 0-5\% most central Pb+Pb collisions at
$\sqrt{s_{NN}}$=2.76 TeV \cite{alice2} as well as the corresponding
results calculated by PACIAE 2.0b, 2.0c, and 2.0c\_c (the same as
PACIAE 2.0c but hadronized by the Monte Carlo coalescence instead of
the string fragmentation).}
\begin{tabular}{cccccc}
\hline\hline
Reaction& $\sqrt{s_{NN}}$ TeV& Exp.& PACIAE 2.0b& PACIAE 2.0c& PACIAE 2.0c\_c\\
\cmidrule[0.25pt](l{0.05cm}r{0.05cm}){1-1}
\cmidrule[0.25pt](l{0.05cm}r{0.05cm}){2-2}
\cmidrule[0.25pt](l{0.05cm}r{0.05cm}){3-3}
\cmidrule[0.25pt](l{0.05cm}r{0.05cm}){4-4}
\cmidrule[0.25pt](l{0.05cm}r{0.05cm}){5-5}
\cmidrule[0.25pt](l{0.05cm}r{0.05cm}){6-6}
Au+Au& 0.2&5060$\pm$250$^\dag$& 4940& 4961& 4746\\
Pb+Pb& 2.76& 1601$\pm$60$^\ddag$& 1554& 1542 & 1540\\
\hline\hline
\multicolumn{6}{l}{$^\dag$ taken from \cite{phob}.
                   $^\ddag$ taken from \cite{alice4}.}\\
\end{tabular}
\label{mul1}
\end{table}
\begin{figure}[ht]
\epsfig{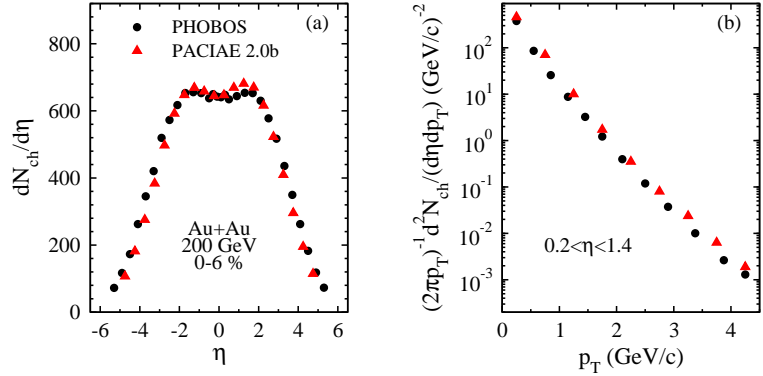}
\caption{(Color online) (a) charged particle pseudorapidity
         distribution in 0-6\% most central Au+Au collisions at $\sqrt{s_{NN}
         }$=200 GeV measured by PHOBOS \cite{phob} and compared with the 
         PACIAE 2.0b results. (b) transverse momentum distribution where data 
         taken from \cite{phob1}).}
\label{auau200}
\end{figure}
   \item The nucleons in one colliding nucleus interact with nucleons
in the other one when they traveling along their straight
trajectories according to the NN total cross section. The
corresponding collision time is calculated for all possible
collision pairs and then the initial NN (hh) collision time list is
composed.
   \item A NN collision pair with least collision time is selected from the
NN (hh) collision time list. This NN (hh) collision, if its cms
energy is large enough ($\sqrt{s}\geq$4.5 GeV, for instance),
executes by PYTHIA with the string fragmentation switched-off and
diquarks broken-up. A parton configuration (parton list) for this NN
(hh) collision is then obtained. If the energy is not large enough
the NN (hh) collision is dealt with the usual two-body elastic
collision \cite{sa1}.
   \item A parton cascade, the same as the one mentioned in last subsection,
proceeds for current NN (hh) collision by parcas\_20b.f.
   \item sfm\_20b.f or coales\_20b.f executes the hadronization of partonic
matter in current NN (hh) collision by the LUND string fragmentation regime 
(calling p20b.f) or Monte Carlo coalescence model according to the switch 
parameter adj1(12). sfm\_20b.f and coales\_20b.f are some what different from 
sfm\_20a.f and coales\_20a.f in PACIAE 2.0a, respectively. p20b.f is 
different from PYTHIA 6.4 in the addition of the mechanism for the reduction 
of strange quark suppression and of the treatment for the charge conservation 
in fragmentation of a string. 
   \item hadcas\_20b.f performs hadron rescattering after hadronization
of current NN (hh) collision. It is similar to the hadron rescattering in
PACIAE 2.0a.
   \item Update the nucleon (hadron) list and NN (hh) collision time list
after current NN (hh) collision executed.
   \item A new NN (hh) collision pair with least collision time is selected
from the NN (hh) collision time list.
   \item Repeat above six items until the NN (hh) collision time list is empty
(hadronic freeze-out). 
   \end{itemize}
  \item A hadronic final state is eventually obtained for an A+B collision
after the execution of parini\_20b.f. 
  \item usu.dat is an example for input file. The left panel of Fig.~
\ref{642} shows the structure of PACIAE 2.0b program for the dynamical 
simulation of a nucleus-nucleus collision.
  \end{enumerate}
 \subsection {PACIAE 2.0c structure}

PACIAE 2.0c program consists of paciae\_20c.f, parini\_20c.f,
parcas\_20c.f, sfm\_20c.f, colaes\_20c.f, hadcas\_20c.f44, and
p20c.f:
\begin{figure}[ht]
\epsfig{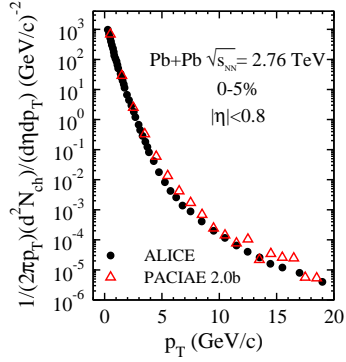}
\caption{(Color online) Charged particle transverse momentum distribution
         in 0-5\% most central Pb+Pb collisions at $\sqrt{s_{NN}}$=2.76 TeV
         measured by ALICE \cite{alice5} and compared with the PACIAE 2.0b
         results.}
\label{pb2760}
\end{figure}
  \begin{enumerate}
  \item paciae\_20c.f plays double roles of the user program (an example) and
the administrating an A+B collision event generation.
  \item parini\_20c.f performs the parton initiation for an A+B collision. It 
is indeed a nucleon cascade process:
   \begin{itemize}
   \item Initiation in the position and momentum spaces for an A+B collision 
is the same as that in PACIAE 2.0b.
   \item The initial NN collision time list is also constructed as the same
as that in PACIAE 2.0b.
   \item A NN collision pair with least collision time is selected from the
NN collision time list.
   \item This NN collision, if energy is large enough, executes by PYTHIA
with the string fragmentation switched-off and the diquarks broken-up.
The generated partons are filled in the parton list. If the energy is not 
large enough the current NN collision is dealt with usual two-body elastic
collision \cite{sa1}.
   \item Update the nucleon list and NN collision time list.
   \item A new NN collision pair with least collision time is selected
from the NN collision time list.
   \item Repeat above three items until the NN collision time list is empty
(hadronic freeze-out). Then one obtains a partonic initial state (parton 
list) for an A+B collision.
   \end{itemize}
  \item parcas\_20c.f performs the parton rescattering for an A+B collision.
  \item sfm\_20c.f executes the hadronization by the LUND string 
fragmentation regime (by calling p20c.f) for an A+B collision if switch 
parameter adj1(12)=0.
  \item coales\_20c.f performs the hadronization for an A+B collision by the
Monte Carlo coalescence model if switch parameter adj1(12)=1.
  \item hadcas\_20c.f executes the hadron rescattering for an A+B collision.
  \item p20c.f is different from PYTHIA 6.4 in the addition of the mechanism 
for reduction of the strange quark suppression and of the requirement for  
charge conservation in the fragmentation of a string.
  \end{enumerate}

The right panel of Fig. \ref{642} shows the structure of PACIAE 2.0c
program for the dynamical simulation of a nucleus-nucleus collision.
Comparing left panel with right panel in Fig. \ref{642}, one knows
that in the PACIAE 2.0b model the parton initiation, parton
rescattering, hadronization, and hadron rescattering are performed
for each hh collision pair independently. This is similar to the
Monte Carlo Glauber model \cite{mcg} despite that it is only a
hadronic cascade model. PACIAE 2.0b characterizes the correlations
among above processes for each hh collision pair, but there are no
correlations among different hh collision pairs. This also means
that in PACIAE 2.0b the nucleus-nucleus collision is dealt as a
superposition of the nucleon-nucleon collisions in a sense. Hence
the PACIAE 2.0b model presents the locality. Oppositely, the PACIAE
2.0c model characterizes the correlations among the different hh
collision pairs in each process of the parton initiation, parton
rescattering, hadronization, and hadron rescattering, but there are
no correlations among above processes for each hh collision pair.
PACIAE 2.0b and 2.0c are similar in the physical contents but are
different in the topological structure.
\begin{figure}[ht]
\epsfig{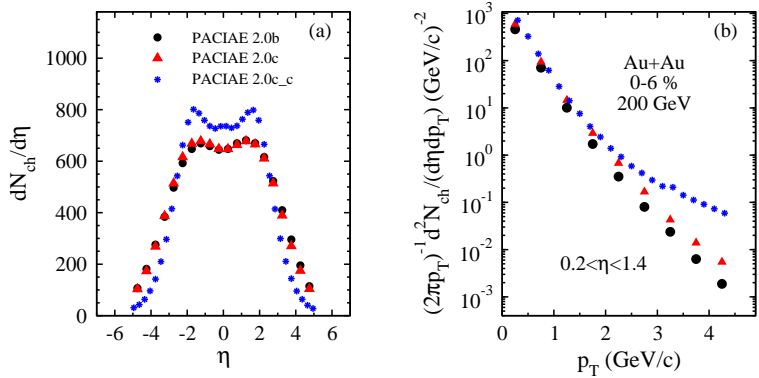}
\caption{(Color online) (a) charged particle pseudorapidity
         distribution in Au+Au collisions at $\sqrt{s_{NN}}$=200 GeV
         calculated by the PACIAE 2.0b (circles), PACIAE 2.0c (triangles),
         and PACIAE 2.0c\_c (stars). (b) transverse momentum distribution.}
\label{auauth}
\end{figure}

In order to decrease the line number in common block `pyjets', a special 
common block `sgam' is introduced. The photon is removed from `pyjets' to 
`sgam' after generation and is given specific flavor code as follows:
\begin{itemize}
\item 22: hardonic decay photon,
\item 44: prompt direct photon,
\item 55: photon from parton-parton rescattering
\begin{equation}
q+g\rightarrow q+\gamma, \hspace{2cm} q+\bar q\rightarrow g+\gamma,
\end{equation}
\item 66: hardonic direct photon
\begin{equation}
\pi+\pi\rightarrow \rho+\gamma, \hspace{2cm} \pi+\rho\rightarrow\pi+\gamma.
\end{equation}
\end{itemize}

In default versions of PACIAE 2.0, the inelastic processes, $q\bar q
\rightarrow gg$ (denoted as process 6 in the program) and $gg\rightarrow 
q\bar q$ (process 7), are switched-off. If one wishes to include the 
inelastic process one has to comment out the statement of `fsqq=0' 
and/or `fsgg\_1=0' in parcas\_20a.f (or parcas\_20b.f, or parcas\_20c.f).
\section{Examples calculated}
The model parameters in PYTHIA 6.4 and PACIAE 2.0 were given default
values according to the experimental measurement and/or the physical
argument. However, in the specific calculations, a few sensitive
parameters, as least as possible, should be tuned to a datum of the
global measurable, such as the charged multiplicity or the charged
particle rapidity density at mid-rapidity. The fitted parameters were
then used to investigate the other physical observables.

In the PACIAE 2.0a calculations the K factor was tuned to the charged
particle pseudorapidity density at mid-pseudorapidity ($|\eta|<1$)
in the INEL (inelastic) pp collisions having at least one charged
particle in the same region (INEL$>0_{|\eta|<1}$) at $\sqrt s$=0.9,
2.36, and 7 TeV \cite{alice1}. The results were shown in
Tab.~\ref{mul} together with the ALICE data \cite{alice1}. This
fitted K=1.9 (D=1.5, D means default value) was employed to calculate
the charged particle pseudorapidity distribution and transverse
momentum distribution ($|\eta|<0.8$) in the INEL pp collisions at
$\sqrt s$=0.9 TeV. These results were compared with the ALICE data
\cite{alice2,alice3} in Fig~\ref{pp900}. One sees in this figure
that the ALICE data are reproduced reasonably.

We have used PACIAE 2.0b to calculate the 0-6\% most central Au+Au
collisions at $\sqrt{s_{NN}}$=200 GeV and 0-5\% most central Pb+Pb
collisions at $\sqrt{s_{NN}}$ =2.76 TeV. In these calculations the K
factor, the parameter $\beta$ in LUND string fragmentation function,
and the time accuracy $\Delta t$ (the least time interval of two
distinguishably consecutive collisions in the parton initiation
stage) were tuned to the PHOBOS data of charged multiplicity
\cite{phob} and the ALICE data of charged particle pseudorapidity
density at mid-rapidity ($|\eta|<0.5$) \cite{alice4}, respectively.
The results were given in Tab. \ref{mul1}. The fitted parameters of K=1.7
(D=1.5), $\beta$=1.5 (D=0.58), and $\Delta t$=0.0001 for Au+Au as well as
K=1.7, $\beta$=1.5, and  $\Delta t$ =0.00004 for Pb+Pb were then employed to
calculate the charged particle pseudorapidity distribution and transverse
momentum distribution in Au+Au collisions and the transverse momentum
distribution in Pb+Pb collisions by PACIAE 2.0b, respectively. These results
were compared with the PHOBOS \cite{phob,phob1} and the ALICE \cite{alice5}
data in Fig.~\ref{auau200} and \ref{pb2760}, respectively. We see in these
figures that the experimental data are well described. The large fluctuation
of the theoretical data at the tail of transverse momentum distribution in
Fig.~\ref{pb2760} is because the total events of 1000 calculated is not
enough.

\begin{figure}[ht]
\epsfig{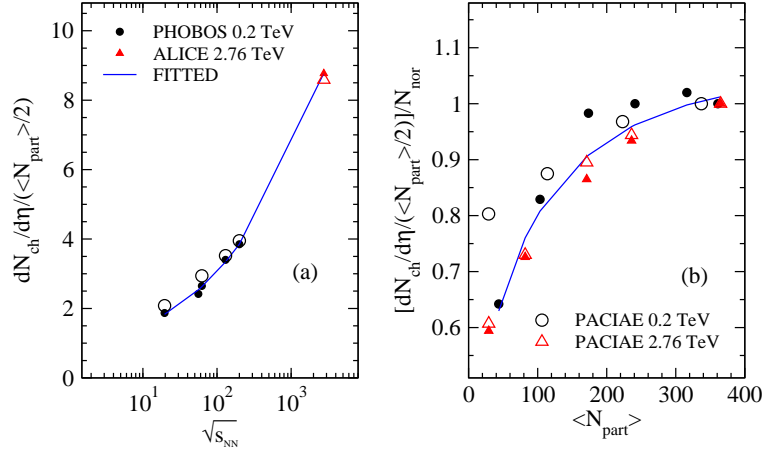}
\caption{(Color online) (a) Energy dependence of the participant scaled
charged particle pseudorapidity density $dN_{ch}/d\eta/(N_{part}/2)$ at
midrapidity in 0-6\% most central Au+Au collisions at RHIC energies and
0-5\% Pb+Pb collisions at LHC energy. (b) Centrality dependence, where
$N_{nor}$ indicates $dN_{ch}/d\eta/(N_{part}/2)$ at largest $N_{part}$.
In this figure the curves are fitted results and the open symbols are
results from PACIAE  2.0b calculations.}
\label{fit}
\end{figure}
Similar calculations were also performed by PAICIA 2.0c and PAICIA 2.0c\_c
(which is the same as PACIAE 2.0c but hadronized by the Mote Carlo
coalescence instead of the string fragmentation ). In the PAICIA 2.0c
calculations K=1.7, $\beta$=0.58, and $\Delta t$=0.0001 were assumed for
Au+Au collisions and K=1.7, $\beta$=0.58, and $\Delta t$=0.00001 for Pb+Pb
collisions. The parameters of K=1.7 and $\Delta t$=0.0001 for Au+Au
collisions as well as K=1.7 and $\Delta t$=0.000055 for Pb+Pb collisions were
assumed in PAICIA 2.0c\_c calculations. The results of charged particle 
multiplicities were compared with the results from PACIAE 2.0b calculations 
and the experimental data in Tab.~\ref{mul1}.  Fig.~\ref{auauth} gave the 
comparison of the charged particle pseudorapidity (panel (a)) and transverse 
momentum (panel (b)) distributions with the corresponding PACIAE 2.0b results 
above. We see in this figure that PAICIA 2.0c is also able to describe the 
experimental data. However, the discrepancy between PAICIA 2.0c and PAICIA 
2.0c\_c is visible.

Recently, PHOBOS investigated the energy dependence of the participant scaled
charged particle pseudorapidity density at mid-rapidity $dN_{ch}/d\eta/
(N_{part}/2)$ in the relativistic pp and A+B collisions. They obtained the 
function of $dN_{ch}/d\eta/(N_{part}/2)$ vs. $\sqrt{s_{NN}}$ by fitting the
experimental data spanned from 10 GeV to 7 TeV for pp and from 2 GeV to 200 
GeV for A+B collisions \cite{phob2} (Before the submission of this paper 
we knew that PHOBOS just published their paper in Phys. Rev. C where they 
added a fitting of the experimental data from 19.6 GeV to 2.76 TeV with a 
power law in $\sqrt{s_{NN}}$ for A+B collisions.). Assuming further that 
the dependence of $dN_{ch}/d\eta/(N_{part}/2)$ on energy and centrality 
(denoted by Glauber model $N_{part}$) can be factorized, then they obtained 
the function of $dN_{ch}/d\eta/(N_{part}/2)$ vs. $\sqrt{s_{NN}}$ and the  
function of $dN_{ch}/d\eta/(N_{part}/2)$ vs. $N_{part}$. 

We used the geometric number of participant nucleons instead of the Glauber 
model $N_{part}$ to fit first the PHOBOS data of 0-6\% most central Au+Au 
collisions at RHIC energies \cite{phob2} and ALICE datum of 0-5\% most 
central Pb+Pb collisions at $\sqrt{s_{NN}}$=2.76 TeV \cite{alice4} (cf.  
Fig.~\ref{fit} (a)). Then we fitted the PHOBOS centrality dependence data
\cite{phob2} with normalization to the datum of highest centrality (i. e. 
$N_{nor}$ in the figure) and ALICE centrality dependence data \cite{alice4} 
with similar normalization (cf. Fig.~\ref{fit}(b)). The fitted functions were 
obtained
\begin{align}
&z=f(x)\frac{g(y)}{g(y_{max})}, \\
&f(x)=0.2\ln(x)+0.001788(\ln(x))^2+0.01244(\ln(x))^3+0.8916, \label{energy} \\
&g(y)=-0.2217+0.3088y^{(1/3)}-0.01906y^{(2/3)}, \label{part}
\end{align}
where $z, x$, and $y$ denoted $dN_{ch}/d\eta/(N_{part}/2)$, $\sqrt{s_{NN}}$,
and $N_{part}$, respectively.

We fitted the PHOBOS pp data \cite{phob2} and ALICE pp data \cite{alice1} of 
$dN_{ch}/d\eta$ by the function as the same as Eq. (\ref{energy}). The new 
four coefficients of 1.725, -0.1312, 0.007777, and -4.3842 were obtained (cf. 
Fig.~\ref{fit_pp}).

With above fitted relations one may be able to first predict $dN_{ch}/d\eta/
(N_{part}/2)$ for un-measured pp and/or A+B collisions in the RHIC and LHC 
energy region. Then tuning the sensitive model parameters (such as K factor, 
$\beta$, and/or $\Delta t$ above) to this $dN_{ch}/d\eta/(N_{part}/2)$ one 
may also be able to predict other observables such as pseudorapidity 
distribution, transverse momentum distribution, etc..
\begin{figure}[ht]
\epsfig{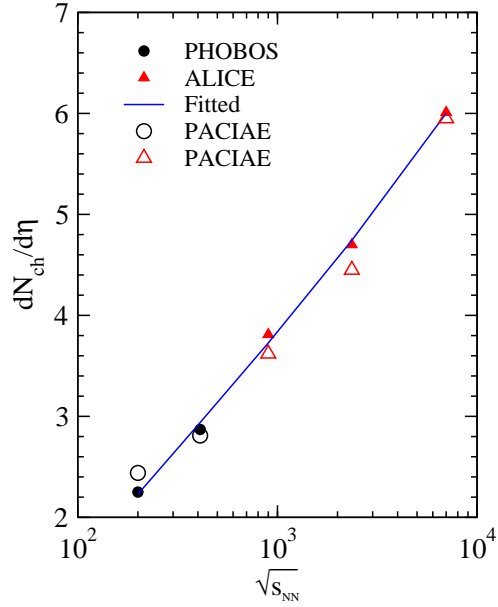}
\caption{(Color online) Energy dependence of the participant scaled charged
particle pseudorapidity density $dN_{ch}/d\eta$ at midrapidity in pp
collisions at RHIC and LHC energies. The curve is fitted results and
the open symbols are results from PACIAE  2.0a calculations.}
\label{fit_pp}
\end{figure}

\begin{flushleft}\bf\large{ACKNOWLEDGEMENT}\end{flushleft}
The financial supports from NSFC (10975062, 11075217, 11047142,10705012) and
from Commission on Higher Education in China are acknowledged. BHS thanks 
Prof. T. Sj\"ostrand for a lot of helps.
\begin{flushleft}\section*\large{\bf APPENDIX: PACIAE 2.0 USERS' GUIDE}\end
{flushleft}
1.  {\bf Selection of the processes}

A switch variable (nchan) is introduced as follows:
 \begin{itemize}
 \item nchan=0: Inelastic (INEL),
 \item nchan=1: Non Single Diffractive (NSD),
 \item nchan=2: $q\bar q\rightarrow \gamma^*/Z^0$ used to generate Drell-Yan
 process,
 \item nchan=3: J/$\psi$ production,
 \item nchan=4: heavy-flavor production,
 \item nchan=5: direct photon.
 \item nchan=6: soft processes only,
 \item nchan=7: default PYTHIA.
 \end{itemize}
2. {\bf Run PACIAE}

 \begin{itemize}
 \item Compile each of the FORTRAN program packets in 20a.tar.gz or 20b.tar.
gz or 20c.tar.gz and link them together forming an executable file.
 \item Set the parameters in input file of usux.dat for PACIAE 2.0a or
usu.dat for PACIAE 2.0b and PACIAE 2.0c properly.
 \item Run the executable file.
 \end{itemize}
3. {\bf Parameters}

The parameters introduced in PACIAE 2.0, except those in PYTHIA 6.4,
can be found in the read statements in user program of paciae\_20a.f
(paciae\_20b.f, paciae\_20c.f). Meaning of parameter can be found in
the program packets via searching its name. Those parameters not
described in the program packets and the array adj1(40) are
explained as follows:
 \begin{itemize}
 \item nout: internal output per `nout' events.
 \item psno=1: systematic sampling method for impact parameter in A+B
               collisions,\\
       psno=2: randomly sampling method for impact parameter.\\
       To run an event with a given impact parameter, set `bmin'='bmax' in
       usu.dat.
 \item adj1(i), i=
 \begin{tabbing}
  \= ttt \= \kill
 \> 1: K factor in parton cascade. \\
 \> 2: $\alpha_s$, effective coupling constant in the parton rescattering
(cascade). \\
 \> 3: parameter `tcut' ($\hat{t}_{cut}$) introduced in the integration
of parton-parton differential \\
 \>\>cross section in the subroutine `fsig' in parcas\_20a.f (parcas\_20b.f,
parcas\_20c.f). \\
 \> 4: parameter `idw', the number of intervals in the numerical integration \\
 \>\> in parcas\_20a.f (parcas\_20b.f, parcas\_20c.f).\\
 \> 5: =1 with nuclear shadowing,\\
 \>\>  =0 without nuclear shadowing.\\
 \> 6: parameter $\alpha$ (parj(41) in PYTHIA 6.4) in the LUND string
fragmentation function. \\
 \> 7: parameter $\beta$ (parj(42) in PYTHIA 6.4) in the LUND string
fragmentation function. \\
 \> 8: mstp(82) in PYTHIA 6.4\\
 \>\>  =1, with hard interactions,\\
 \>\>  =0: without hard interactions (simple two-string model, soft only).\\
 \> 9: parp(81) in PYTHIA 6.4 (default=1.4 GeV/c), effective minimum
transverse momentum for\\
 \>\> multiple interactions when mstp(82)=1.\\
 \> 10: K factor (parp(31) in PYTHIA 6.4). \\
 \> 11: time accuracy used in the hadron cascade (hadcas\_20a.f, hadcas\_20b.f, and hadcas\_20c.f). \\
 \> 12: model for hadronization \\
 \> \> =0 string fragmentation,\\
 \> \> =1 Monte Carlo coalescence model. \\
 \> 13: dimension of meson table considered.\\
 \> 14: dimension of baryon table considered.\\
 \> 15: string tension.\\
 \> 16: number of loops in the deexcitation of energetic quark in the Monte
Carlo
coalescence model.\\
 \> 17: the threshold energy in the deexcitation of energetic quark in the
Monte Carlo coalescence model.\\
 \> 18: =0 without Pauli blocking in the parton cascade, \\
 \>\>   =1 with Pauli blocking.\\
 \> 19: time accuracy used in the parton cascade (parcas\_20a.f, parcas
\_20b.f, and parcas\_20c.f). \\
 \> 20: =0 using LO pQCD parton-parton cross section in the parton
rescattering (cascade),\\
 \>\>    =1 using keeping only leading divergent terms in the LO pQCD
parton-parton cross section,\\
 \>\>    =2 the same as 0 but flat scattering angle distribution is assumed,\\
 \>\>    =3 the same as 1 but flat scattering angle distribution is assumed.\\
 \> 21: =0 without phase space constraint in the Monte Carlo coalescence model,
\\
 \>\>   =1 with phase space constraint.\\
 \> 22: critical value of the product of radii in the position and momentum
phase spaces (4 is assumed).\\
 \> 23: =0 LUND string fragmentation function is used in the subroutine for
deexcitation of the \\
 \>\>  energetic quark in the Monte Carlo coalescence model,\\
 \>\> =1 Field-Feynman parametrization function is used.\\
 \> 24: the virtuality cut ('tl0') in the time-like radiation in the parton
rescattering (cascade). \\
 \> 25: $\Lambda_{QCD}$ in the parton cascade. \\
 \> 26: number of random number thrown away. \\
 \> 27: largest momentum allowed for particle.\\
 \> 28: concerned to the largest position allowed for particle, see program 
        for the detail.\\
 \> 29: width of two dimension Gaussian distribution sampling $p_x$
        and $p_y$ of the produced quark \\
 \>\>   pair in the deexcitation of energetic quark in the Monte Carlo
coalescence model.
\\
 \> 30: maximum $p_T^2$ in above two dimension Gaussian distribution.\\
 \> 31: parj(1) in PYTHIA 6.4. \\
 \> 32: parj(2) in PYTHIA 6.4. \\
 \> 33: parj(3) in PYTHIA 6.4. \\
 \> 34: parj(21) in PYTHIA 6.4. \\
 \> 35: mstp(91) in PYTHIA 6.4, parton transverse momentum ($k_{\perp}$)
distribution inside hadron;\\
 \>\>   =1, Gaussian;\\
 \>\>   =2, exponential\\
 \> 36: =0 without phenomenological parton energy loss in the parton
rescattering (cascade), \\
 \>\>   =1 with phenomenological parton energy loss. \\
 \> 37: the coefficient in phenomenological parton energy loss.\\
 \> 38: $p_T$ cut in phenomenological parton energy loss.\\
 \> 39: width of Gaussian $k_{\perp}$ distribution in hadron if mstp(91)=1,\\
 \>\>   width of exponential $k_{\perp}$ distribution in hadron if mstp(91)=2.
    \\
 \> 40: decide where the run stopped:\\
 \>\>   =1, after parton initiation;\\
 \>\>   =2, after parton rescattering;\\
 \>\>   =4, after hadron rescattering. \\
 \end{tabbing}
 \end{itemize}
4. {\bf Output files}

 \begin{itemize}
 \item rms0.out is an output for the input parameters.
 \item rms.out is the main output file.
 \item main.out is the PYTHIA standard output.
 \item oscar.out is the OSCAR standard output if nosc=2 or 3.
 \item Encourage user to write his/her own output file and user program.
 \end{itemize}
5. {\bf Postscript}

 \begin{itemize}
 \item The PACIAE 2.0 program is free for the person who is interested.
However, it is protected by GPL, no merchandise use please.
 \item Welcome reporting any bugs to yanyl@ciae.ac.cn, zhoudm@phy.ccnu.edu.cn
, and/or sabh@ciae.ac.cn .
 \end{itemize}
\begin{flushleft}\bf\large{REFERENCES}\end{flushleft}

\newpage
}
\end{document}